\documentclass{PoS}

\usepackage{amsmath}

\usepackage{calrsfs}
\usepackage[vcentermath]{youngtab}
\usepackage{ytableau}

\newcommand{\Useries}[4]{\foreach \index in {1,...,#2}{#1_{#3_\index}^{\ 
#4_\index}}}
\usepackage{tikz}
\usetikzlibrary{matrix,arrows,decorations.pathmorphing,decorations.markings,patterns}
\tikzstyle arrowstyle=[scale=1]
\tikzstyle directed=[postaction={decorate,decoration={markings,
    mark=at position .65 with {\arrow[arrowstyle]{stealth}}}}]
\tikzstyle reverse directed=[postaction={decorate,decoration={markings,
    mark=at position .65 with {\arrowreversed[arrowstyle]{stealth};}}}]
\newcommand{\EQ}[1]{\begin{equation} #1 \end{equation}}

\newcommand{\SP}[1]{\begin{equation}\begin{split} #1 \end{split}\end{equation}}

\newcommand*{\tr}[0]{{\rm tr}}

\newcommand{\field}[1]{\mathbb{#1}}

\title{Calculating the chiral condensate diagrammatically at strong coupling}

\ShortTitle{$\langle {\bar \psi} \psi \rangle$ vs. $N_f$ at $g = \infty$}

\author{Alexander S. Christensen\\
        Niels Bohr Institute, Blegdamsvej 17, 2100 Copenhagen, Denmark\\
        E-mail: \email{xander@nbi.dk}}

\author{\speaker{Joyce C. Myers}\\
        Niels Bohr Institute, Blegdamsvej 17, 2100 Copenhagen, Denmark\\
        E-mail: \email{jcmyers@nbi.dk}}
        
\author{Peter D. Pedersen\\
        Niels Bohr Institute, Blegdamsvej 17, 2100 Copenhagen, Denmark\\
        E-mail: \email{peter.pedersen@nbi.dk}}
        
\author{Jan Rosseel\\
        Vienna University of Technology, Wiedner Hauptstr., 8-10/136, A-1040 Vienna, Austria\\
        E-mail: \email{rosseelj@hep.itp.tuwien.ac.at}}

\abstract{We calculate the chiral condensate of QCD at infinite coupling as a function of the number of fundamental fermion flavours using a lattice diagrammatic approach inspired by recent work of Tomboulis, and other work from the 80's. We outline the approach where the diagrams are formed by combining a truncated number of sub-diagram types in all possible ways. Our results show evidence of convergence and agreement with simulation results at small $N_f$. However, contrary to recent simulation results, we do not observe a transition at a critical value of $N_f$. We further present preliminary results for the chiral condensate of QCD with symmetric or adjoint representation fermions as a function of $N_f$ for $N_c = 3$. In general, there are sources of error in this approach associated with miscounting of overlapping diagrams, and over-counting of diagrams due to symmetries. These are further elaborated upon in a longer paper.}

\FullConference{The 32nd International Symposium on Lattice Field Theory,\\
		23-28 June, 2014\\
		Columbia University New York, NY}

\begin{document}

\section{Introduction}

Lattice diagrammatic techniques can be valuable tools to obtain insight into the strong coupling limit of QCD and related theories. We consider a particular diagrammatic approach which was introduced in the 80's to study chiral symmetry breaking in QCD at infinite coupling, as $N_f \rightarrow 0$, in \cite{Blairon:1980pk}, and then further developed in \cite{Martin:1982tb}. More recently this approach has been picked up again to address the question of chiral symmetry restoration in the case of QCD with a large number of fermion flavours $N_f$. In particular, the simulation results in \cite{deForcrand:2012vh} for the chiral condensate at infinite coupling as a function of $N_f$ show evidence of a first order transition to a chiral symmetry restored phase at a critical value of $N_f \sim 13$  staggered flavours. Although such a transition is well documented at more moderate coupling strengths, its presence at infinite coupling came as a surprise, because analytical calculations based on a $1/d$ expansion \cite{KlubergStern:1982bs}, or mean field \cite{Damgaard:1985bn}, suggested that chiral symmetry would remain broken for all $N_f$ at infinite coupling. The lattice diagrammatic technique of \cite{Blairon:1980pk,Martin:1982tb} was then reintroduced and extended to account for contributions arising at nonzero $N_f$ in \cite{Tomboulis:2012nr}. There are two solutions for the normalised chiral condensate as a function of $N_f$ obtained in \cite{Tomboulis:2012nr}. One of these solutions matches onto \cite{Martin:1982tb} in the $N_f \rightarrow 0$ limit, where the normalised chiral condensate goes to $\sim 0.66$ as $N_f \rightarrow 0$, then increases in magnitude as $N_f$ increases. The other solution goes to infinity as $N_f \rightarrow 0$, and decreases as a function of $N_f$. For both solutions, there is a common critical value of $N_f \sim 10.7$, beyond which only complex solutions for the chiral condensate exist. It would be good to understand this better. The idea of this note, and of our recent longer paper in \cite{Christensen:2014dia}, is to develop a procedure, inspired by \cite{Tomboulis:2012nr}, which can be used to calculate the chiral condensate by collecting the contributions from all possible diagrams which can be formed out of a truncated number of sub-diagram types.

\section{$\langle {\bar \psi} \psi \rangle$ at $g = \infty$}

As in \cite{Tomboulis:2012nr}, we begin by generalising the procedure in \cite{Martin:1982tb} to incorporate contributions which arise at nonzero $N_f$. The iterative procedure we employ to generalise \cite{Martin:1982tb} is different from that of \cite{Tomboulis:2012nr}, and we summarise it below using the notation of \cite{Martin:1982tb,Tomboulis:2012nr}.

Integrating out the fermion fields puts the chiral condensate in the form
\EQ{
\langle {\bar \psi}(x) \psi(x) \rangle = - \lim_{m\rightarrow 0} \tr \left[ \frac{\int dU \, \det\left[ 1 + K^{-1} M(U) \right] \left[ \left[ 1 + K^{-1} M(U) \right]^{-1} K^{-1} \right]_{xx}}{\int dU \, \det\left[ 1 + K^{-1} M(U) \right]} \right] \, ,
\label{2ptcorr}
}
with
\EQ{
M_{xy} \equiv \frac{1}{2} \sum_\mu \left[ \gamma_{\mu} U_{\mu}(x) \delta_{y,x+{\hat \mu}} - \gamma_{\mu} U_{\mu}^{\dag}(x-{\hat \mu}) \delta_{y,x-{\hat \mu}} \right] \, , \hspace{1cm} K_{xy} = m {\field I}_{N_f} {\field I}_{N_c} \delta_{x y} \, .
}
for $\mu = 1, ..., d$. The form of $\langle {\bar \psi} \psi \rangle$ in (\ref{2ptcorr}) suggests expanding in powers of $K^{-1} M$, resulting in
\EQ{
\det\left[ 1 + K^{-1} M \right] = \exp \tr \left[ \sum_{n=1}^{\infty} \frac{(-1)^{n+1}}{n} (K^{-1} M)^n \right] \, ,
\label{detD}
}
\EQ{
\left[ \left[ 1 + K^{-1} M \right]^{-1} K^{-1} \right]_{xx} = \frac{1}{m} \left[ \sum_{n=0}^{\infty} (-1)^n (K^{-1}M)^n \right]_{xx} \, .
}
The presence of the trace in (\ref{detD}) and (\ref{2ptcorr}) allows for simplifications using $\tr \left[ \text{odd \# of} ~\gamma_{\mu}\text{'s} \right] = 0$, such that only contributions from terms with $(K^{-1}M)^n$ for $n$ even are nonzero. In addition, due to the $SU(N_c)$ integrals over the $U$'s, the only nonzero diagrams are those where each link has $U^m U^{\dag n}$, for some $m$, $n$, such that $m-n \equiv 0 \hspace{-1mm} \mod 
N_c$.

Following \cite{Martin:1982tb} the normalised chiral condensate can be put in the form
\EQ{
\frac{1}{N_s N_f N_c} \langle {\bar \psi} \psi \rangle = - \lim_{m \rightarrow 0} ~ \frac{1}{m} \sum_{L=0}^{\infty} (-1)^L 
\frac{A(L)}{(2 m)^{2L}} \, ,
}
where $A(L)$ is the contributions from all graphs with $2 L$ links which start and end at some site $x$. A general graph can be built out of irreducible graphs with less links (if the graph is not already irreducible). Specifically, an irreducible graph cannot be separated into smaller graphs which start and end at $x$.

\vspace{-2mm}
\begin{minipage}{0.5\textwidth}

\hspace{3cm}
\begin{tikzpicture}[scale=0.5]

\def \xoff {10.0}
\def \yoff {0.0}

\draw [directed] (0.0+\xoff,0.0+\yoff) -- (0.0+\xoff,1.8+\yoff);
\draw [directed] (0.0+\xoff,1.8+\yoff) -- (-1.0+\xoff,3.2+\yoff) -- 
(-0.8+\xoff,3.4+\yoff) -- (0.2+\xoff,1.9+\yoff) -- (0.4+\xoff,3.8+\yoff) -- 
(0.6+\xoff,3.7+\yoff) -- (0.4+\xoff,2.0+\yoff) -- (2.1+\xoff,2.8+\yoff) -- 
(2.3+\xoff,2.6+\yoff) -- (0.2+\xoff,1.7+\yoff);
\draw [directed] (0.2+\xoff,1.7+\yoff) -- (0.2+\xoff,0.0+\yoff);

\node[anchor=west] at (\xoff-0.5,-0.5) {\Large{
$x$
}};

\node[anchor=west] at (\xoff-2.0,5.0) {\large{
Irreducible
}};

\end{tikzpicture}\\
\end{minipage}
\begin{minipage}{0.5\textwidth}

\hspace{0cm}
\begin{tikzpicture}[scale=0.5]

\def \xoff {14.0}

\draw (0.0+\xoff,0.0) -- (-1.0+\xoff,1.4) -- (-0.8+\xoff,1.6) -- 
(0.2+\xoff,0.25);
\draw [directed] (0.2+\xoff,0.25) -- (2.1+\xoff,1.0) -- (2.3+\xoff,2.9) -- 
(2.5+\xoff,2.8) -- (2.3+\xoff,0.8) -- (0.2+\xoff,0.0);

\node[anchor=west] at (\xoff-0.5,-0.5) {\Large{
$x$
}};

\node[anchor=west] at (\xoff-1.7,4.0) {\large{
Reducible
}};

\end{tikzpicture}
\end{minipage}\vspace{-3mm}

To obtain the contribution of all general diagrams $A(L)$ with $2 L$ links, it is necessary to take all possible combinations of irreducible graphs $I(l)$ of $2 l$ links, which form a diagram of $2 L$ links,
\EQ{
A(L) = \sum_{l=1}^{L} I(l) A(L-l) \, , \hspace{1cm} L \ge 1 \, ; \hspace{1cm} A(0) = 1 \, ,\vspace{-3mm}
}
where the irreducible graphs can begin with an area-$0$ contribution, a)
\begin{tikzpicture}[scale=0.3]

\draw [directed] (0.0,0.0) -- (0.0,2.0);
\draw (0.0,2.0) -- (0.2,2.0);
\draw [directed] (0.2,2.0) -- (0.2,0.0);

\end{tikzpicture}
, or an area $1$ base diagram, such as b)
\begin{tikzpicture}[scale=0.3]

\def \xoff {0.0}
\def \yoff {0.0}

\draw [directed] (0.0+\xoff,0.0+\yoff) -- (0.0+\xoff,2.0+\yoff) -- (2.0+\xoff,2.0+\yoff) -- (2.0+\xoff,0.0+\yoff) -- (0.2+\xoff,0.0+\yoff);
\draw [reverse directed] (0.2+\xoff,0.2+\yoff) rectangle (1.8+\xoff,1.8+\yoff);

\end{tikzpicture}
, or ... . The first four $I(l)$ are
\vspace{-2mm}
\begin{flalign}
\begin{tikzpicture}[baseline=(current  bounding  box.center),scale=0.5]
\node[anchor=east] at (0.2,1.0) {\large{
$I(1) = ~$
}};
\draw [directed] (0.0,0.0) -- (0.0,2.0);
\draw (0.0,2.0) -- (0.2,2.0);
\draw [directed] (0.2,2.0) -- (0.2,0.0);
\node[anchor=west] at (0.2,1.0) {\large{
$= I_a(1) = 2 d$\ ,
}};
\def \xoff {13.0}
\node[anchor=east] at (0.2+\xoff,1.0) {\large{
$I(2) = ~$
}};
\draw [directed] (0.0+\xoff,0.0) -- (0.0+\xoff,2.0);
\draw [directed] (0.0+\xoff,2.0) -- (2.0+\xoff,2.7) -- (2.1+\xoff,2.5) -- (0.2+\xoff,1.8);
\draw [directed] (0.2+\xoff,1.8) -- (0.2+\xoff,0.0);
\node[anchor=west] at (0.2+\xoff,1.0) {\large{
$= I_a(2) = 2 d ~  \left[ I_a(1) ~ {\widehat a}_0 \right]$\ ,
}}; 
\end{tikzpicture}&&
\end{flalign}
\vspace{-7mm}
\begin{flalign}
\begin{tikzpicture}[baseline=(current  bounding  box.center),scale=0.5]
\def \xoff {6.0}
\node[anchor=east] at (0.2,1.0) {\large{
$I(3) = ~$
}};
\draw [directed] (0.0,0.0) -- (0.0,2.0);
\draw [directed] (0.0,2.0) -- (1.9,2.7) -- (1.5,4.5) -- (1.7,4.5) -- (2.2,2.5) 
-- (0.2,1.8);
\draw [directed] (0.2,1.8) -- (0.2,0.0);
\node[anchor=west] at (0.2,1.0) {\large{
$\hspace{10mm}+$
}};
\draw [directed] (0.0+\xoff,0.0) -- (0.0+\xoff,1.8);
\draw [directed] (0.0+\xoff,1.8) -- (-1.5+\xoff,3.2) -- (-1.4+\xoff,3.4) -- 
(0.1+\xoff,2.0) -- (1.9+\xoff,2.7) -- (2.1+\xoff,2.5) -- (0.2+\xoff,1.8);
\draw [directed] (0.2+\xoff,1.8) -- (0.2+\xoff,0.0);
\node[anchor=west] at (0.2+\xoff,1.0) {\large{
$~~= I_a(3) = 2 d \left[ I_a(2) {\widehat a}_0 + I_a(1)^2 {\widehat a}_0^2 
\right]$\ ,
}}; 
\end{tikzpicture}&&
\end{flalign}
\vspace{-6mm}
\begin{flalign}
\begin{tikzpicture}[baseline=(current  bounding  box.center),scale=0.5]
\def \xoff {5.0}
\node[anchor=east] at (0.2,1.0) {\large{
$I(4) = ~$
}};
\draw [directed] (0.0,0.0) -- (0.0,2.0);
\draw [directed] (0.0,2.0) -- (1.9,2.7) -- (1.5,4.5) --(3.0,6.2) -- (3.2,6.0) -- 
(1.8,4.5) -- (2.2,2.5) -- (0.2,1.8);
\draw [directed] (0.2,1.8) -- (0.2,0.0);
\node[anchor=west] at (0.2,1.0) {\large{
$\hspace{10mm}+$
}};
\draw [directed] (0.0+\xoff,0.0) -- (0.0+\xoff,1.8);
\draw [directed] (0.0+\xoff,1.8) -- (1.9+\xoff,2.7) -- (0.5+\xoff,4.2) -- 
(0.6+\xoff,4.4) -- (2.1+\xoff,2.8) -- (3.0+\xoff,4.7) -- (3.2+\xoff,4.6) -- 
(2.2+\xoff,2.6) -- (0.2+\xoff,1.7);
\draw [directed] (0.2+\xoff,1.7) -- (0.2+\xoff,0.0);
\def \xoff {10.5}
\node[anchor=west] at (0.2+5.0,1.0) {\large{
$\hspace{10mm}+ ~~~~2$
}};
\draw [directed] (0.0+\xoff,0.0) -- (0.0+\xoff,1.8);
\draw [directed] (0.0+\xoff,1.8) -- (-1.0+\xoff,3.2) -- (-0.8+\xoff,3.4) -- 
(0.2+\xoff,1.9) -- (2.1+\xoff,2.8) -- (2.3+\xoff,4.7) -- (2.5+\xoff,4.6) -- 
(2.3+\xoff,2.6) -- (0.2+\xoff,1.7);
\draw [directed] (0.2+\xoff,1.7) -- (0.2+\xoff,0.0);
\def \xoff {16.0}
\def \yoff {0.0}
\node[anchor=west] at (0.2+10.0,1.0+\yoff) {\large{
$\hspace{15mm}+$
}};
\draw [directed] (0.0+\xoff,0.0+\yoff) -- (0.0+\xoff,1.8+\yoff);
\draw [directed] (0.0+\xoff,1.8+\yoff) -- (-1.0+\xoff,3.2+\yoff) -- 
(-0.8+\xoff,3.4+\yoff) -- (0.2+\xoff,1.9+\yoff) -- (0.4+\xoff,3.8+\yoff) -- 
(0.6+\xoff,3.7+\yoff) -- (0.4+\xoff,2.0+\yoff) -- (2.1+\xoff,2.8+\yoff) -- 
(2.3+\xoff,2.6+\yoff) -- (0.2+\xoff,1.7+\yoff);
\draw [directed] (0.2+\xoff,1.7+\yoff) -- (0.2+\xoff,0.0+\yoff);
\def \xoff {20.0}
\def \yoff {0.0}
\node[anchor=west] at (0.2+15.5,1.0+\yoff) {\large{
$\hspace{10mm}+$
}};
\draw [directed] (0.0+\xoff,0.0+\yoff) -- (0.0+\xoff,2.0+\yoff) -- 
(2.0+\xoff,2.0+\yoff) -- (2.0+\xoff,0.0+\yoff) -- (0.2+\xoff,0.0+\yoff);
\draw [reverse directed] (0.2+\xoff,0.2+\yoff) rectangle (1.8+\xoff,1.8+\yoff);
\node[anchor=west] at (0.2-2.5,-1.0) {\large{
$~~= I_a(4)+ I_b(4) ~~= 2 d \left[ I_a(3) {\widehat a}_0 + 2 I_a(1) I_a(2) {\widehat a}_0^2 + 
I_a(1)^3 {\widehat a}_0^3 \right] - 4d(d-1) \frac{N_f}{N_c}$\ ,
}};
\end{tikzpicture}&&
\end{flalign}
\vspace{-8mm}
\begin{flalign}
\begin{tikzpicture}[baseline=(current  bounding  box.center),scale=0.5]
\def \xoff {6.0}
\node[anchor=east] at (0.2,1.0) {\large{
$...$\ .
}};
\end{tikzpicture}&&
\end{flalign}
We have defined $I_a(l)$ as all irreducible graphs of length $2 l$ starting with a)
\begin{tikzpicture}[scale=0.3]

\draw [directed] (0.0,0.0) -- (0.0,2.0);
\draw (0.0,2.0) -- (0.2,2.0);
\draw [directed] (0.2,2.0) -- (0.2,0.0);

\end{tikzpicture}
, $I_b(l)$ as all irreducible graphs of length $2 l$ starting with b)
\begin{tikzpicture}[scale=0.3]

\def \xoff {0.0}
\def \yoff {0.0}

\draw [directed] (0.0+\xoff,0.0+\yoff) -- (0.0+\xoff,2.0+\yoff) -- (2.0+\xoff,2.0+\yoff) -- (2.0+\xoff,0.0+\yoff) -- (0.2+\xoff,0.0+\yoff);
\draw [reverse directed] (0.2+\xoff,0.2+\yoff) rectangle (1.8+\xoff,1.8+\yoff);

\end{tikzpicture}
, etc. The ${\widehat x}_n$ are defined as ${\widehat x}_n \equiv \frac{x_n}{d_x}$, where $x_n$ is the number of ways of attaching a type $x$ diagram to an area $n$ diagram, defined to reduce over-counting, and $d_x$ is the total dimensionality of a type $x$ diagram. For example, ${\widehat a}_0 = \frac{2d-1}{2d}$, ${\widehat b}_0 = \frac{4(d-1)^2}{4d(d-1)}$. More ${\widehat x}_n$ are defined in appendix A of \cite{Christensen:2014dia}.
In general the $I(l)$ can thus be put in the form
\SP{
I(l) = &2 d F_0(l-1) - 4 d(d-1) \frac{N_f}{N_c} F_1(l-4)^7 + ... \,, \hspace{1cm} \text{with}~ I(0) = 0 \, ,
\label{IL}
}
where $F_n(L)$ represents all possible graphs of length $2 L$ which start and end on a site on a base diagram of area $n$. The $F_n$ are composed of all possible combinations of irreducible graphs which add up to $2 L$ links,
\EQ{
F_n(L) = \sum_{\substack{l_i = 1, 2, ... ,\\ k_j = 4, 8, ...,\\ \sum l_i + k_j = 
L - 1}} I_a(l_1) I_a(l_2) ... I_a(l_p) I_b(k_1) I_b(k_2) ... I_b(k_q) ... ~ 
\widehat{a}_n^p\, \widehat{b}_n^q ... \, , \hspace{9mm} \text{with}~ F_n(0) = 1 \, .
\label{Fn}
}

The generating function for all irreducible graphs, including the mass dependence, is
\EQ{
W_I = \sum_{l=0}^{\infty} \left( -\frac{1}{4 m^2} \right)^l I(l) = W_a + W_b + ... \, , \vspace{-2mm}
}
where $W_a$ is all irreducible graphs starting with an $a$-type base diagram
\begin{tikzpicture}[scale=0.3]

\draw [directed] (0.0,0.0) -- (0.0,2.0);
\draw (0.0,2.0) -- (0.2,2.0);
\draw [directed] (0.2,2.0) -- (0.2,0.0);

\end{tikzpicture}
, $W_b$ is all irreducible graphs starting with a $b$-type base diagram
\begin{tikzpicture}[scale=0.3]

\def \xoff {0.0}
\def \yoff {0.0}

\draw [directed] (0.0+\xoff,0.0+\yoff) -- (0.0+\xoff,2.0+\yoff) -- 
(2.0+\xoff,2.0+\yoff) -- (2.0+\xoff,0.0+\yoff) -- (0.2+\xoff,0.0+\yoff);
\draw [reverse directed] (0.2+\xoff,0.2+\yoff) rectangle (1.8+\xoff,1.8+\yoff);

\end{tikzpicture}
, etc. Using (\ref{Fn}) and (\ref{IL}) gives
\EQ{
W_a = 2 d x \sum_{n=0}^{\infty} \left[ {\widehat a}_0 W_a + {\widehat b}_0 W_b + 
... \right]^n = \frac{2 d x}{1 - {\widehat a}_0 W_a - {\widehat b}_0 W_b - ...} 
\, ,
\label{Wa}
}
\EQ{
W_b = -4d(d-1) \frac{N_f}{N_c} x^4 \left[ \sum_{n=0}^{\infty} \left[ {\widehat 
a}_1 W_a + {\widehat b}_1 W_b + ...\right]^n \right]^7 = 
\frac{-4d(d-1)\frac{N_f}{N_c} x^4}{(1 - {\widehat a}_1 W_a - {\widehat b}_1 W_b 
- ... )^7} \, ,
}
\EQ{
... \, ,
\label{Who}
}
where $x \equiv - \frac{1}{4 m^2}$ and the ``$...$" contain irreducible graphs starting with higher order (in $\frac{1}{m}$) base diagrams. The chiral condensate is obtained by taking all possible combinations of all possible irreducible diagrams. That is
\EQ{
\frac{\langle \bar{\psi} \psi \rangle}{N_s N_f N_c} = \lim_{m \rightarrow 0} 
\frac{1}{m} \left( \frac{1}{1-W_I} \right) \, .
\label{W-chicon}
}
It is possible to obtain a simpler system of equations than (\ref{Wa}) - (\ref{Who}) by working in the massless limit. One can introduce the variables $g_x \equiv - \frac{2 m W_x}{d_x}$, such that, taking $m \rightarrow 0$,
\EQ{
g_a = \frac{1}{a_0 g_a + b_0 g_b + ...} \, , \hspace{5mm} g_b = \frac{\frac{N_f}{N_c}}{(a_1 g_a + b_1 g_b + ...)^7} \, , \hspace{5mm} g_c = \frac{\frac{N_f}{N_c}}{(a_2 g_a + b_2 g_b + ...)^{11}} \, , \hspace{5mm} ... \, .
\label{gall}
}
The chiral condensate can then be obtained from $g \equiv d_a g_a + d_b g_b + ...$, using
\EQ{
\frac{\langle \bar{\psi} \psi \rangle}{N_s N_f N_c} = \frac{2}{g} \, .
\label{chicong}
}
The prefactors in the numerators of (\ref{gall}), and the powers of the quantity in the denominators need to be determined for each diagram type. The total contribution of a diagram includes
\begin{itemize}
\item A factor $\frac{1}{i!} (-N_f N_s)^i$, for a number $i$, of overlapping 
closed internal loops,
\item A mass factor $\left( - \frac{1}{4m^2} \right)^n$, for $n$ pairs of links,
\item $(-1)^k$ for $k$ permutations of $\gamma$ matrices,
\item A factor containing the result obtained by performing the 
group integrations,
\item A factor containing the dimensionality of the graph. 
\end{itemize}
Group integrals for overlapping links of the form
\begin{tikzpicture}[scale=0.3]

\draw [directed] (0.0,0.0) -- (0.0,2.0);
\draw [directed] (0.4,2.0) -- (0.4,0.0);

\end{tikzpicture}
, or
\begin{tikzpicture}[scale=0.3]

\draw [directed] (0.0,0.0) -- (0.0,2.0);
\draw [directed] (0.4,2.0) -- (0.4,0.0);
\draw [directed] (0.8,0.0) -- (0.8,2.0);
\draw [directed] (1.2,2.0) -- (1.2,0.0);

\end{tikzpicture}
are nonzero $\forall \, N_c$, given by \cite{Wilson:1975id,Bars:1979xb,Creutz:1984mg,Cvitanovic:2008zz}
\EQ{
\int_{{\rm SU}(N_c)} d U \, U_{a}^{\ b} (U^\dag)_{c}^{\ d} = \frac{1}{N_c} 
\delta_{a}^{d} \delta_{c}^{b} \, ,
}
\SP{
\int_{{\rm SU}(N_c)} d U \, \Useries{U}{2}{a}{b} \Useries{(U^\dag)}{2}{c}{d} = 
&\tfrac{1}{2 N_c(N_c+1)} \left( \delta_{a_1}^{d_1} \delta_{a_2}^{d_2} + 
\delta_{a_1}^{d_2} \delta_{a_2}^{d_1} \right) \left( \delta_{c_1}^{b_1} 
\delta_{c_2}^{b_2} + \delta_{c_1}^{b_2} \delta_{c_2}^{b_1} \right) \\
&+ \tfrac{1}{2 N_c(N_c-1)} \left(  \delta_{a_1}^{d_1} 
\delta_{a_2}^{d_2} - \delta_{a_1}^{d_2} \delta_{a_2}^{d_1}  \right) 
\left(\delta_{c_1}^{b_1} \delta_{c_2}^{b_2} - \delta_{c_1}^{b_2} 
\delta_{c_2}^{b_1}  \right) \, .
}
For finite $N_c$, for example
\begin{tikzpicture}[scale=0.3]

\draw [directed] (0.0,0.0) -- (0.0,2.0);
\draw [directed] (0.4,0.0) -- (0.4,2.0);
\draw [directed] (0.8,0.0) -- (0.8,2.0);

\end{tikzpicture}
for $SU(3)$, integrals of the form
\begin{equation}
\int_{{\rm SU}(N_c)} d U\, U_{a_1}^{\ b_1} \cdots U_{a_{N_c}}^{\ b_{N_c}} = 
\frac{1}{N_c!}\epsilon_{a_1 \cdots  a_{N_c}} \epsilon^{b_1\cdots b_{N_c}} \,.
\end{equation}
are needed. These rules are sufficient to evaluate the diagrams we will use, including
\begin{flalign}
\begin{tikzpicture}[baseline=(current  bounding  box.center),scale=0.7]
\node[anchor=west] at (0.0,1.0) {\large{$a$)}};
\def \xoff {1.2}
\draw [directed] (0.0+\xoff,0.0) -- (0.0+\xoff,2.0);
\draw (0.0+\xoff,2.0) -- (0.2+\xoff,2.0);
\draw [directed] (0.2+\xoff,2.0) -- (0.2+\xoff,0.0);
\def \xoff {3.2}
\def \yoff {0}
\node[anchor=west] at (2.0,1.0) {\large{$b$)}};
\draw [directed] (0.0+\xoff,0.0) -- (0.0+\xoff,2.0) -- (2.0+\xoff,2.0) -- (2.0+\xoff,0.0) -- (0.2+\xoff,0.0);
\draw [reverse directed] (0.2+\xoff,0.2) rectangle (1.8+\xoff,1.8);
\def \xoff {7.2}
\def \yoff {0}
\node[anchor=west] at (6.0,1.0) {\large{$d$)}};
\draw [directed] (0.0+\xoff,0.0) -- (0.0+\xoff,2.0) -- (2.0+\xoff,2.0) -- (2.0+\xoff,0.0) -- (0.2+\xoff,0.0);
\draw [directed] (0.2+\xoff,0.2) rectangle (1.8+\xoff,1.8);
\draw [directed] (0.4+\xoff,0.4) rectangle (1.6+\xoff,1.6);
\def \xoff {10.2}
\def \yoff {0}
\draw [directed] (0.0+\xoff,0.0) -- (0.0+\xoff,2.0) -- (2.0+\xoff,2.0) -- (2.0+\xoff,0.0) -- (0.2+\xoff,0.0) 
-- (0.2+\xoff,1.8) -- (1.8+\xoff,1.8) -- (1.8+\xoff,0.2) -- (-0.2+\xoff,0.2);
\draw [directed] (0.4+\xoff,0.4) rectangle (1.6+\xoff,1.6);
\def \xoff {13.2}
\def \yoff {0}
\draw [directed] (0.0+\xoff,0.0) -- (0.0+\xoff,2.0) -- (2.0+\xoff,2.0) -- (2.0+\xoff,0.0) -- (0.2+\xoff,0.0)
                             -- (0.2+\xoff,1.8) -- (1.8+\xoff,1.8) -- (1.8+\xoff,0.2) -- (0.4+\xoff,0.2)
                             -- (0.4+\xoff,1.6) -- (1.6+\xoff,1.6) -- (1.6+\xoff,0.4) -- (-0.4+\xoff,0.4);
\def \xoff {16.2}
\def \yoff {0}
\draw [directed] (0.0+\xoff,0.0) -- (0.0+\xoff,2.0) -- (2.0+\xoff,2.0) -- (2.0+\xoff,0.0) -- (0.2+\xoff,0.0);
\draw [directed] (0.2+\xoff,0.4) -- (0.2+\xoff,1.8) -- (1.8+\xoff,1.8) -- (1.8+\xoff,0.2) -- (0.4+\xoff,0.2) 
-- (0.4+\xoff,1.6) -- (1.6+\xoff,1.6) -- (1.6+\xoff,0.4) -- (0.2+\xoff,0.4);
\def \xoff {1.0}
\def \yoff {-3.0}
\node[anchor=west] at (0.0,-2.0) {\large{$g$)}};
\draw [directed] (0.0+\xoff,0.0+\yoff) -- (0.0+\xoff,2.0+\yoff) -- (2.0+\xoff,2.0+\yoff) -- (2.0+\xoff,0.0+\yoff) -- (0.2+\xoff,0.0+\yoff);
\draw [reverse directed] (0.2+\xoff,0.2+\yoff) rectangle (1.8+\xoff,1.8+\yoff);
\draw [directed] (0.4+\xoff,0.4+\yoff) rectangle (1.6+\xoff,1.6+\yoff);
\draw [reverse directed] (0.6+\xoff,0.6+\yoff) rectangle (1.4+\xoff,1.4+\yoff);
\def \xoff {4.0}
\def \yoff {-3.0}
\draw [directed] (0.0+\xoff,0.0+\yoff) -- (0.0+\xoff,2.0+\yoff) -- (2.0+\xoff,2.0+\yoff) -- (2.0+\xoff,0.0+\yoff) -- (0.2+\xoff,0.0+\yoff);
\draw [reverse directed] (0.2+\xoff,0.4+\yoff) -- (0.2+\xoff,1.8+\yoff) -- (1.8+\xoff,1.8+\yoff) -- (1.8+\xoff,0.2+\yoff) -- (0.4+\xoff,0.2+\yoff) -- (0.4+\xoff,1.6+\yoff) -- (1.6+\xoff,1.6+\yoff) -- (1.6+\xoff,0.4+\yoff) -- (0.2+\xoff,0.4+\yoff);
\draw [directed] (0.6+\xoff,0.6+\yoff) rectangle (1.4+\xoff,1.4+\yoff);
\def \xoff {7.0}
\def \yoff {-3.0}
\draw [directed] (0.0+\xoff,0.0+\yoff) -- (0.0+\xoff,2.0+\yoff) -- (2.0+\xoff,2.0+\yoff) -- (2.0+\xoff,0.0+\yoff) -- (0.2+\xoff,0.0+\yoff) 
-- (0.2+\xoff,1.8+\yoff) -- (1.8+\xoff,1.8+\yoff) -- (1.8+\xoff,0.2+\yoff) -- (-0.2+\xoff,0.2+\yoff);
\draw [reverse directed] (0.4+\xoff,0.6+\yoff) -- (0.4+\xoff,1.6+\yoff) -- (1.6+\xoff,1.6+\yoff) -- (1.6+\xoff,0.4+\yoff) -- 
(0.6+\xoff,0.4+\yoff) -- (0.6+\xoff,1.4+\yoff) -- (1.4+\xoff,1.4+\yoff) -- (1.4+\xoff,0.6+\yoff) -- (0.4+\xoff,0.6+\yoff);
\node[anchor=west] at (2.4+\xoff,-2.0) {\large{.}};
\end{tikzpicture}&&
\label{type-all}
\end{flalign}
The specific contributions of these (and other) diagrams are given in \cite{Christensen:2014dia}.
\section{Group integration with Young Projectors}

To calculate higher order diagrams one needs to evaluate integrals of the general form
\EQ{
I_n \equiv \int_{SU(N_c)} {\rm d}U~ U_{\alpha_1}{}^{\beta_1} ... U_{\alpha_n}{}^{\beta_n} (U^{\dag})_{\gamma_1}{}^{\delta_1} ... (U^{\dag})_{\gamma_n}{}^{\delta_n}
\label{In}
}
Any nonzero integral including some combination of $U$, $U^{\dag}$ can be converted to this form using $U_{a_1}^{\ b_1} = \frac{1}{(N-1)!} \epsilon_{a_1 a_2 \cdots a_N} \epsilon^{b_1 b_2 \cdots b_N} (U^\dag)_{b_2}^{\ a_2} \cdots (U^\dag)_{b_N}^{\ a_N}$ and $(U^\dag)_{a_1}^{\ b_1} = \frac{1}{(N-1)!} \epsilon_{a_1 a_2 \cdots a_N} \epsilon^{b_1 b_2 \cdots b_N} U_{b_2}^{\ a_2} \cdots U_{b_N}^{\ a_N}$. Calculating the direct product of $n$ $U$'s ($U^{\dag}$'s) leads to a direct sum of representations $R$ ($S$). The integral can be obtained from the Young Projectors ${\field P}$ of these representations using \cite{Cvitanovic:2008zz}
\EQ{
\int_{SU(N_c)} {\rm d}U~ R_{a}{}^b (S^{\dag})_c{}^d = \frac{1}{d_R} ({\field P}^{R})_a{}^d ({\field P}^{S})_{c}{}^b ~\delta_{R S} \, .
}

Consider for example the integral in (\ref{In}) with $n = 2$. The direct product ${\bf N_c} \otimes {\bf N_c}$ is
\EQ{
\ytableausetup{aligntableaux=center}
\ytableaushort[\alpha_]{1} \otimes \ytableaushort[\alpha_]{2} = \ytableaushort[\alpha_]{12} \oplus \ytableaushort[\alpha_]{1,2}\,.
}
The Young projectors are thus formed by symmetrising, and antisymmetrising in $\alpha_1$ and $\alpha_2$,
\EQ{
\mathbb{P}^S_{\alpha_1 \alpha_2}{}^{\beta_1\beta_2}  =  \frac12 \left(\delta_{\alpha_1}^{\beta_1} \delta_{\alpha_2}^{\beta_2} + \delta_{\alpha_1}^{\beta_2} \delta_{\alpha_2}^{\beta_1}   \right) \,, \hspace{10mm} \mathbb{P}^{AS}_{\alpha_1 \alpha_2}{}^{\beta_1\beta_2}  =  \frac12 \left(\delta_{\alpha_1}^{\beta_1} \delta_{\alpha_2}^{\beta_2} - \delta_{\alpha_1}^{\beta_2} \delta_{\alpha_2}^{\beta_1}   \right) \,.
}
The resulting integral is
\EQ{
I_2 = \frac{2}{N_c(N_c+1)} \mathbb{P}^S_{\alpha_1 \alpha_2}{}^{\delta_1 \delta_2} \mathbb{P}^S_{\gamma_1 \gamma_2}{}^{\beta_1 \beta_2} + \frac{2}{N_c(N_c-1)} \mathbb{P}^{AS}_{\alpha_1 \alpha_2}{}^{\delta_1 \delta_2} \mathbb{P}^{AS}_{\gamma_1 \gamma_2}{}^{\beta_1 \beta_2}\,.
}
More involved examples ($I_3$, $I_4$) are worked out in \cite{Christensen:2014dia}.

\section{Higher dimensional representations}

Higher dimensional representations can be written in terms of the fundamental and anti-fundamental. For example, the symmetric $(U^S)_{a}{}^{b}$, for $a, b = 1, ..., d_{S}$, is given by
\SP{
(U^S)_{(\alpha_1 \alpha_2)}{}^{(\beta_1 \beta_2)} &= ({\field P}^{S})_{\alpha_1 \alpha_2}{}^{\gamma_1 \gamma_2} U_{\gamma_1}{}^{\delta_1} U_{\gamma_2}{}^{\delta_2} ({\field P}^{S}) _{\delta_1 \delta_2}{}^{\beta_1 \beta_2} = \frac{1}{2} \left( U_{\alpha_1}{}^{\beta_1} U_{\alpha_2}{}^{\beta_2} + U_{\alpha_1}{}^{\beta_2} U_{\alpha_2}{}^{\beta_1} \right) \, .
}
The antisymmetric $(U^{AS})_{m}{}^{n}$, for $m, n = 1, ..., d_{AS}$, is given by
\SP{
(U^{AS})_{[\alpha_1 \alpha_2]}{}^{[\beta_1 \beta_2]} &= ({\field P}^{AS})_{\alpha_1 \alpha_2}{}^{\gamma_1 \gamma_2} U_{\gamma_1}{}^{\delta_1} U_{\gamma_2}{}^{\delta_2} ({\field P}^{AS}) _{\delta_1 \delta_2}{}^{\beta_1 \beta_2} = \frac{1}{2} \left( U_{\alpha_1}{}^{\beta_1} U_{\alpha_2}{}^{\beta_2} - U_{\alpha_1}{}^{\beta_2} U_{\alpha_2}{}^{\beta_1} \right) \, .
}
The adjoint $(U^{A})_{a}{}^{b}$, for $a, b = 1, ..., d_{Adj}$, can be written as
\EQ{
(U^{A})_{a}{}^{b} = 2 \, \mathrm{Tr}\left( U t_a U^\dag t^b\right)\,,
}
where the  $t_a$ are fundamental generators of $SU(N_c)$ normalised as $\mathrm{Tr}\left(t_a t_b\right) = \frac12 \delta_{ab}$. For integrals with higher dimensional representation links in the form
\begin{tikzpicture}[scale=0.3]

\draw [directed] (0.0,0.0) -- (0.0,2.0);
\draw [directed] (0.4,2.0) -- (0.4,0.0);

\end{tikzpicture}
, it is sufficient to use
\EQ{
\int_{SU(N)} {\rm d}U~ (U^{R})_{a}{}^{b} (U^{R \dag})_{c}{}^{d} = \frac{1}{d_R} \delta_{a}{}^{d} \delta_{c}{}^{b} \, . 
}
Further considering the adjoint, we are in general interested in integrals with links of the form
\begin{tikzpicture}[scale=0.3]

\draw [directed] (0.0,0.0) -- (0.0,2.0);
\draw [directed] (0.4,0.0) -- (0.4,2.0);
\draw [directed] (1.6,0.0) -- (1.6,2.0);

\node[anchor=west] at (0.02,0.8) {\large{...}};

\end{tikzpicture},
for $n$ lines, that is
\SP{
I^{A}_n &\equiv \int \, d U \, U_{a_1}{}^{b_1} \cdots U_{a_n}{}^{b_n} \\
&= 2^n\,  (t_{a_1})_{\beta_1}{}^{\gamma_1} (t^{b_1})_{\delta_1}{}^{\alpha_1} \cdots (t_{a_n})_{\beta_n}{}^{\gamma_n} (t_{b_n})_{\delta_n}{}^{\alpha_n} \int \, d U\, U_{\alpha_1}{}^{\beta_1} \cdots U_{\alpha_n}{}^{\beta_n} U^{\dag}_{\gamma_1}{}^{\delta_1} \cdots U^{\dag}_{\gamma_n}{}^{\delta_n}
}
For example, for $n=3$, evaluating the fundamental integral and simplifying using the identity $t_a t_b = \frac{1}{2N} \delta_{ab} \mathbf{1}_N + \frac12 d_{abc} t_c + \frac{i}{2} f_{abc} t_c$, results in
\EQ{
I^{A}_3 = \frac{N_c}{(N_c^2-1)(N_c^2-4)} d_{a_1 a_2 a_3}d^{b_1 b_2 b_3} + \frac{1}{N_c(N_c^2-1)} f_{a_1 a_2 a_3}f^{b_1 b_2 b_3} \,.
}
where $i f_{abc}= 2 \, \mathrm{Tr}\left([t_a,t_b]t_c \right)$, $d_{abc}= 2 \, \mathrm{Tr}\left( \{t_a,t_b\} t_c \right)$.

\section{Results}

\begin{figure}[t]
\begin{minipage}{0.5\textwidth}
\includegraphics[scale=0.53]{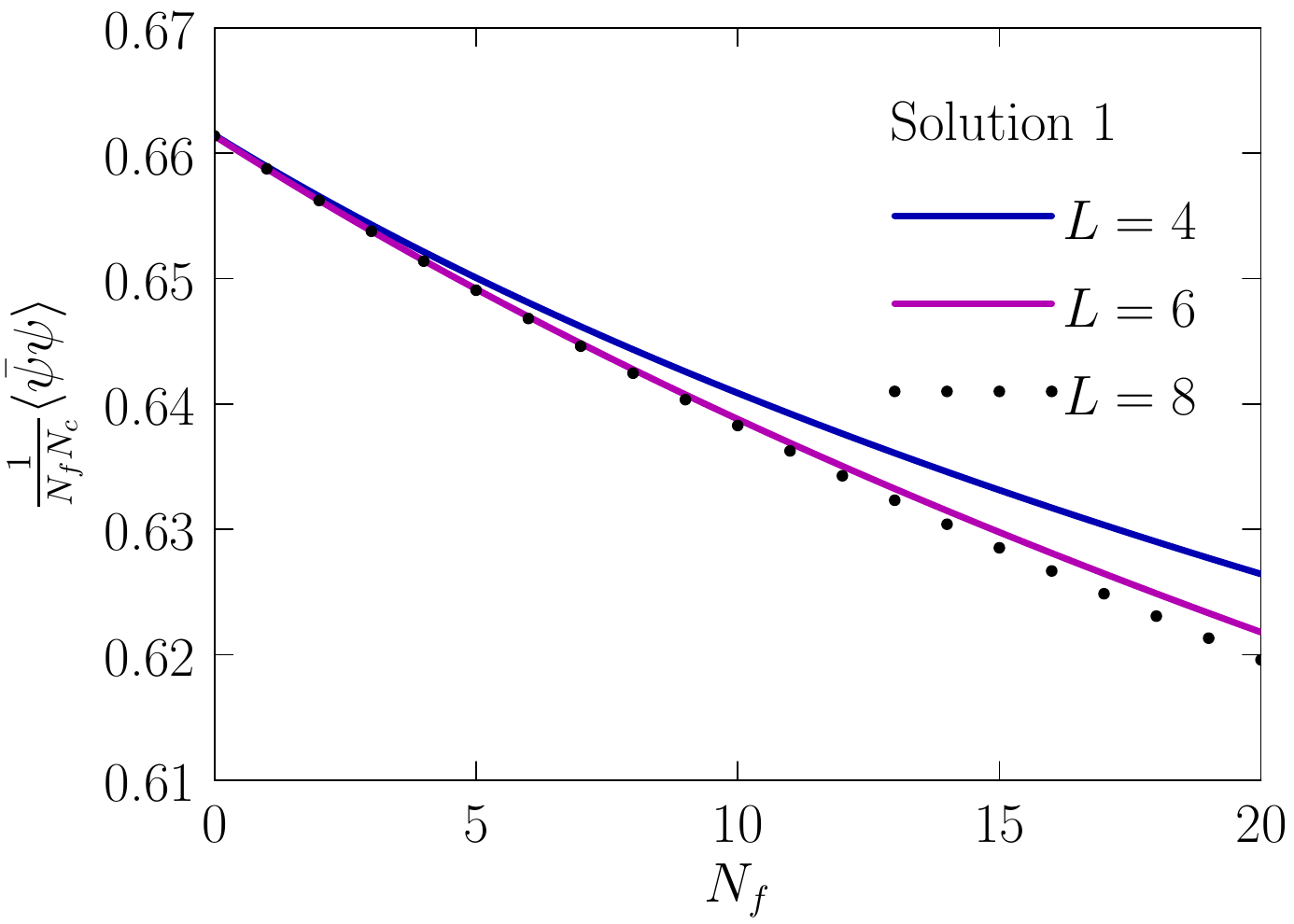}
\end{minipage}
\begin{minipage}{0.5\textwidth}
\includegraphics[scale=0.53]{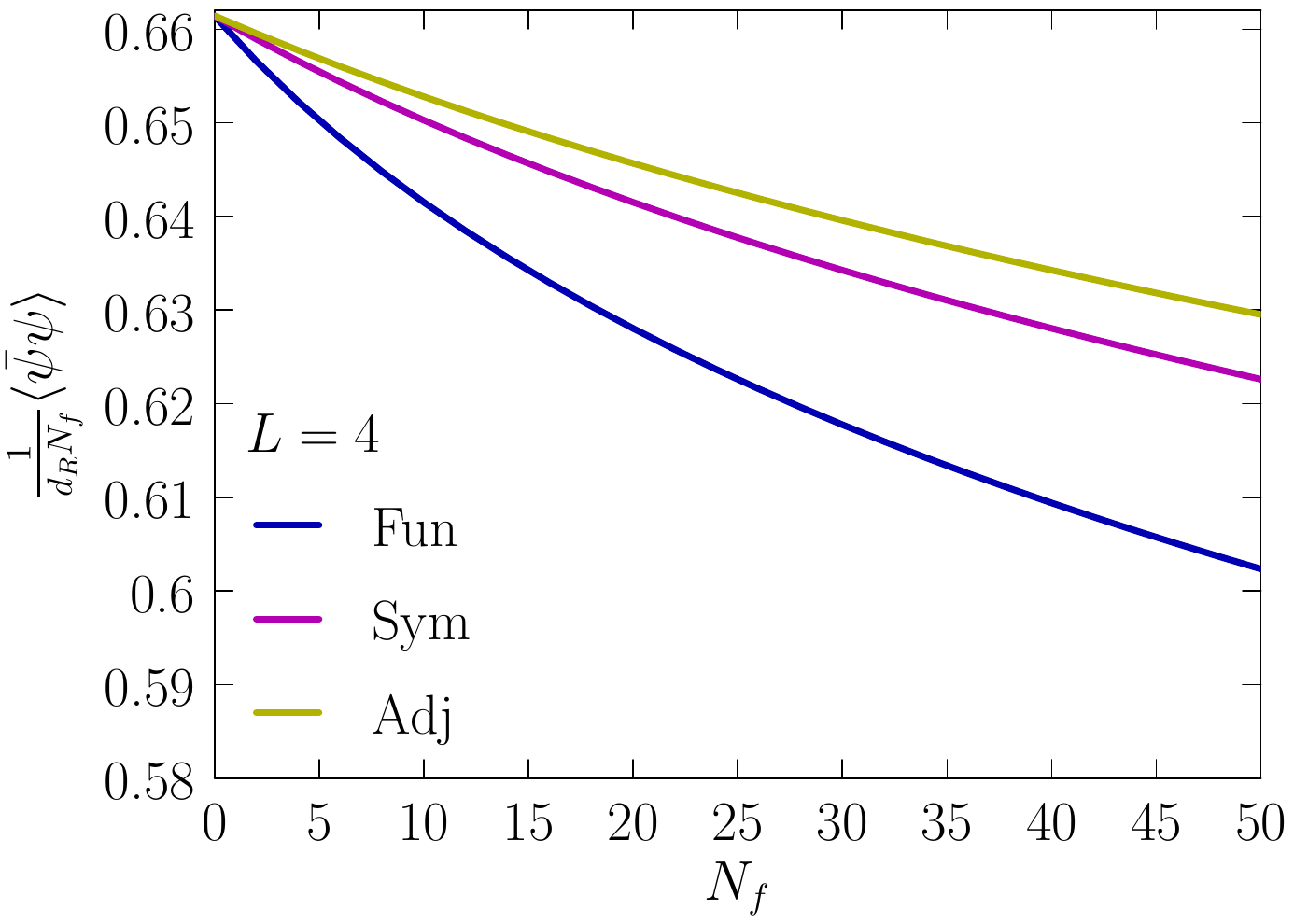}
\end{minipage}
\caption{$\frac{1}{N_f d_R} \langle {\bar \psi} \psi \rangle$ vs. $N_f$ at $g = \infty$ including area $1$ sub-diagrams up to order $\left(\frac{1}{m^2}\right)^L$ for $L = 4, 6, 8$ with fermions in the fundamental representation (left), and comparing the fundamental, symmetric, and adjoint, including sub-diagrams up to $L = 4$ (right).}
\label{sols}
\end{figure}

Results for the normalised chiral condensate $\frac{1}{N_f d_R} \langle {\bar \psi} \psi \rangle$ are plotted in Figure \ref{sols}. The solution plotted is that which goes to the result of \cite{Martin:1982tb,deForcrand:2006gu} in the $N_f \rightarrow 0$ limit. A more detailed analysis of results is presented in \cite{Christensen:2014dia}. A remarkable feature of these results is that as $N_f$ is increased, the chiral condensate decreases very slowly and approaches zero as $N_f \rightarrow \infty$. Unlike in \cite{deForcrand:2012vh,Tomboulis:2012nr}, there is no indication of discontinuity in any of the solutions obtained. However, we cannot rule out that the preferred solution changes at some critical $N_f$. There are sources of error associated with this approach including mis-counting of overlapping diagrams, and over-counting due to symmetries. These need to be quantified. For details see \cite{Christensen:2014dia}.

\end{document}